\documentclass[aps,floats,epsf,color]{revtex4}
\usepackage{graphicx}
\usepackage{amssymb}
\begin{document}
\def\be{\begin{equation}}
\def\ee{\end{equation}}
\def\bea{\begin{eqnarray}}
\def\eea{\end{eqnarray}}
\def\E{{\rm e}}
\def\bearst{\begin{eqnarray*}}
\def\eearst{\end{eqnarray*}}
\def\peleven{\parbox{11cm}}
\def\peffec{\peight{\bearst\eearst}\hfill\peleven}
\def\pspace{\peight{\bearst\eearst}\hfill}
\def\ptwelve{\parbox{12cm}}
\def\peight{\parbox{8mm}}


\title
{Discretely Holomorphic Parafermions in Lattice Z$_N$ Models}

\author
{M.A.~Rajabpour$^{a,b}$\footnote{e-mail:
rajabpour@physics.sharif.edu}, J.~Cardy$^{b,c}$\footnote{e-mail:
j.cardy1@physics.ox.ac.uk} }

\address
{$^{a}$Department of Physics, Sharif University of Technology,
Tehran, Iran 11365-9161\\$^{b}$Rudolf Peierls Centre for
Theoretical Physics, 1 Keble Road, Oxford OX1 3NP, UK\\ $^{c}$All
Souls College, Oxford}
\date{\today}

\begin{abstract}
We construct lattice parafermions - local products of order and
disorder operators - in nearest-neighbour Z$_N$ models on regular
isotropic planar lattices, and show that they are discretely
holomorphic, that is they satisfy discrete Cauchy-Riemann
equations, precisely at the critical Fateev-Zamolodchikov (FZ)
integrable points. We generalise our analysis to models with
anisotropic interactions, showing that, as long as the lattice is
correctly embedded in the plane, such discretely holomorphic
parafermions exist for particular values of the couplings which we
identify as the anisotropic FZ points. These results extend to
more general inhomogeneous lattice models as long as the covering
lattice admits a rhombic embedding in the plane.

\end{abstract}
\maketitle
\section{Introduction}
Z$_N$ models, as the simplest statistical mechanics models with
discrete global symmetries, have been studied for many years. They
are interesting, both as lattice models and in the continuum limit
as quantum field theories, because they exhibit semi-locality
properties, such as fractional spin \cite{Fradkin-Kadanoff}, which
arise in various other domains of physics. The simplest examples
of these models are the well-known Ising model ($N=2$),
three-state Potts model ($N=3$), and Ashkin-Teller model ($N=4$),
all exactly solvable to some extent. The phase diagram of the
nearest neighbour Z$_N$ model in two dimensions for $N=5,6,7$ was
elucidated in \cite{alcaraz koberle,cardyZn}, but in the general
case it is still not known. Although these lattice models for
$N>4$ have complicated phase diagrams with several critical
surfaces, Fateev and Zamolodchikov \cite{FZ star triangle} showed
that there are some points (hereinafter referred to as FZ points)
in the critical surface at which these models are solvable, in the
sense that they satisfy generalised star-triangle relations.
Moreover there is strong evidence that the scaling limit at these
points corresponds to Zamolodchikov and Fateev's parafermionic
conformal field theory \cite{parafermion FZ}. In this article we
give further evidence for this connection. In particular we find
lattice candidates for the holomorphic parafermions of the
continuum model, and show that these satisfy a discrete version of
the Cauchy-Riemann equations. We also extend this result to models
with anisotropic interactions. We show that as long as the lattice
is correctly embedded in the plane, we can again construct
discretely holomorphic parafermions. We may use this to conjecture
the location of the FZ critical surface of these anisotropic Z$_N$
lattice models. In fact, our methods can be extended to any
non-uniform lattice of the Baxter type \cite{baxter}, thanks to a
result of Kenyon and Schlenker \cite{kenyon} about embedding such
lattices in the plane.

Identifying discretely holomorphic objects in lattice models is an
important step in showing that the scaling limit of suitably
defined curves in these models is described by Schramm-Loewner
evolution (SLE) \cite{smirnov,cardy}. Suitable candidates for SLE
curves in Z$_N$ models have been suggested by Santachiara
\cite{santachiara} (see also Gamsa and Cardy \cite{gamsacardy} for
the case $N=3$). We expect that our results will be the first step
in showing that this is indeed the case.

This paper is organised as follows. In the second section we
briefly review some related properties of Z$_N$ models, and the
corresponding parafermionic conformal field theories. We
 define the disorder variables on the isotropic
square lattice, similarly to the case of the Z$_N$ clock models
\cite{Fradkin-Kadanoff} and the Potts model \cite{knops}. We then
define lattice parafermions as products of neighbouring order and
disorder variables, with a suitable phase factor. Using the
Boltzmann weights of the model at the FZ point, we then show that
there are local linear relations between parafermions, which are
equivalent to discrete holomorphicity. This is our first main
result.

In the third section we consider lattices with anisotropic
interactions and show that there are special points, which we
identify with anisotropic FZ points, at which there is a certain
embedding of the lattice on the plane for which discrete
holomorphicity is again recovered. We further extend this to the
case of Baxter lattices \cite{baxter}, where we show that the
existence of a rhombic embedding \cite{kenyon} for such lattices
allows us simply to generalise our results. It should be mentioned
that the importance of rhombic embeddings (and the related
isoradial embeddings) was central to the work of Duffin
\cite{duffin}, Mercat \cite{mercat} and Bobenko \em et al.\em\
\cite{bobenko} in the general theory of discrete holomorphy.
Similar ideas have been used in other lattice models by Kenyon
\cite{kenyon2} and Bazhanov \em et al. \em\cite{bazhanov}.

\section{Holomorphic Parafermions in Z(N) Model}
\label{sec2} The general Z$_N$ model on an arbitrary graph can be
defined by associating to every node of the graph a variable
$s_{r}$ which takes values in the set $\omega^{q}$,
$q=0,1,...,N-1$, where $\omega=e^{{2\pi i}/{N}}$. Each
configuration occurs with probability
\begin{eqnarray}\label{partition function}
Z^{-1}\prod_{(rr')}W(r,r')\,,
\end{eqnarray}
where the product is over all the edges $(rr')$ of the graph, and
$Z$ is the partition function. The weights $W(r,r')$ take the
general form
\begin{eqnarray}
W(r,r')=\sum_{k=0}^{N-1}x^{(rr')}_{k}\big(s_{r}s^{*}_{r'}\big)^{k}\,,
\end{eqnarray}
where $^*$ denotes complex conjugation. Assuming
$x^{(rr')}_0\not=0$, we can always set it equal to unity. Reality
of the weights implies that $x^{(rr')}_{k}=x^{(rr')}_{N-k}$, which
means that one can describe them by just $[{N}/{2}]$ real
parameters. The weights are invariant under the global group
Z$_{N}$, $s_{r}\rightarrow \omega^{k}s_{r}$, as well as charge
conjugation $C: s_r\to s_r^*$.

In what follows, except for the last section, we consider the
graph to be a regular lattice, with translationally invariant
couplings $x_k^{(rr')}$, which may however depend on the lattice
orientation of the edge $(rr')$.

For $N=2$ and $3$ it is easy to see that the model is equivalent
to the Ising model and three state Potts model respectively, whose
critical points correspond to well understood minimal conformal
field theories with $c<1$. The higher values of $N$ are related to
more complicated theories with rich phase diagrams\cite{alcaraz}.
Nearest-neighbour Z$_N$ models exhibit Kramers-Wannier duality
symmetry, which is important in understanding their phase
diagrams. If all of the $x_{k}$ with $k\geq 1$ have the same
value, we have the well-known $N$-state Potts model, which has a
single first-order transition for $N>4$. In general for $N>3$ the
theory has a critical surface on which the exponents may vary
continuously. It has been conjectured that there are $n<\frac12N$
points in the phase diagram which correspond to special kind of
conformal field theories called Fateev-Zamolodchikov (FZ)
parafermionic models of the first kind \cite{parafermion FZ}. This
has been verified by numerical simulations \cite{alcaraz}. Our
lattice holomorphic operators, described below, are another
evidence for this conjecture. These conformal field theories have
the central charge
\begin{eqnarray}\label{central charge}
c=\frac{2(N-1)}{N+2}\,,
\end{eqnarray}
 the cases $N=2,3$ coinciding with the
central charge of $p=2,5$ minimal CFTs. Kramers-Wannier symmetry
manifests itself in $N-1$ order parameter fields $\sigma_{m}$ and
$N-1$ dual disorder fields $\mu_m$, which are conjectured to be
the continuum version of the order and disorder variables on the
original lattice. Zamolodchikov and Fateev, using only the Z$_{N}$
symmetry, showed that there also exist holomorphic operators (that
is, operators whose correlation functions are analytic functions
except at coincident points) with the following dimensions (equal
to their conformal spin)
\begin{eqnarray}\label{holomorphic dimensions}
p_{m}=\frac{m(N-m)}{N},\hspace{2cm}m=1,...,N-1\,.
\end{eqnarray}

We now give our construction of the lattice analogs of these
objects. We first consider the homogenous isotropic Z$_N$ model on
the square lattice.

In the first step let define the disorder operators
$\mu_{\tilde{r}}$ on the sites of dual lattice, which, in this
case, is again another square lattice whose vertices are at the
centers $\tilde{r}$ of the faces of the original lattice. The
insertion of a disorder operator $\mu_{\tilde{r}m}$ corresponds to
modifying the weights so that the order operator $s_{r}$ has
monodromy $s_{r}\rightarrow \omega^{-m}s_{r}$ on taking the point
$r$ in a closed circuit around $\tilde{r}$. This is equivalent to
introducing a path, or string, on the sites of the dual lattice
from $\tilde{r}$ to infinity (or some other point on the
boundary), such that the weights on edges $(rr')$ intersected by
the string are modified by the substitution $s_rs^*_{r'}\to
s_r\omega^{-m}s^*_{r'}$.

Thus the disorder operator has the following form for the general
Z$_N$ model
\begin{eqnarray}\label{disorder operator}
\mu_{\tilde{r}m}=\prod_{\mbox{$(rr')$ intersected by string}}
\frac{\sum_{k=0}^{N-1}x_{k}(s_{r}\omega^{-m}s^{*}_{r'})^{k}}{\sum_{k=0}^{N-1}x_{k}(s_{r}s^{*}_{r'})^{k}}\,.
\end{eqnarray}

It is not difficult to see that the disorder variables have the
same $C$-symmetry similar to the spin variables,
$\mu_{m}=\mu_{N-m}$, and that, up to a gauge transformation, the
definition (\ref{disorder operator}) is path-independent.



\begin{figure}[h]
\centering
\includegraphics[width=6cm]{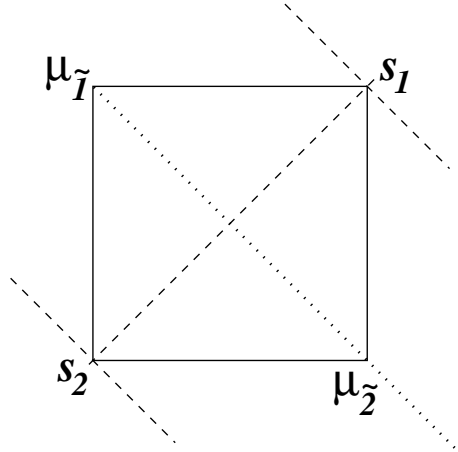}
\caption{\label{fig1} An elementary square of the covering lattice
(solid lines -- rotated by 45$^\circ$ to correspond to the
conventions in the text.) The opposite pairs of vertices are
associated with order ($s$) and disorder ($\mu$) operators
respectively. The parafermions $\psi$ are associated with the
edges. Discrete holomorphicity at the FZ points means that the
contour sum of the parafermions around each face vanishes. Also
shown is part of the original square lattice (dashed lines) and
the string (dotted line) attached to the disorder operators. }
\end{figure}

Consider the square which the vertices are made by the two
neighbouring spin variables $s_{1},s_{2}$ and the two neighbouring
disorder variables $\mu_{\tilde{1}},\mu_{\tilde{2}}$
(Fig1.~\ref{fig1}). By taking the string from $\tilde1$ to run
along the dual edge $(\tilde1\tilde2)$, from (\ref{disorder
operator}) we can write the following relation
\begin{eqnarray}\label{relation between disorder operators}
\mu_{\tilde{1}m}=
\frac{\sum_{k=0}^{N-1}x_{k}(s_{1}\omega^{-m}s^{*}_{2})^{k}}{\sum_{k=0}^{N-1}x_{k}(s_{1}s^{*}_{2})^{k}}
\mu_{\tilde{2}m}\,.
\end{eqnarray}
We will need the exact value of the $x_{k}$ at the FZ point, which
for the isotropic square lattice have the following compact form
\cite{FZ star triangle}
\begin{eqnarray}\label{FZpoint}
x_{ck}=\prod_{j=0}^{k-1}\frac{\sin(\frac{\pi
j}{N}+\frac{\pi}{4N})}{\sin(\frac{\pi
(j+1)}{N}-\frac{\pi}{4N})}\,.
\end{eqnarray}

For clarity, let us first focus on the simplest case $N=2$. By
multiplying both side of (\ref{relation between disorder
operators}) by the denominator and then by multiplying the result
by $s_{1}$ we find the following equation
\begin{eqnarray}\label{s equation}
s_{1}\mu_{\tilde{1}}+x_{c}s_{2}\mu_{\tilde{1}}=s_{1}\mu_{\tilde{2}}-x_{c}s_{2}\mu_{\tilde{2}}\,.
\end{eqnarray}
A similar equation can be found by exchanging
$s_{1}\leftrightarrow s_{2}$. By combining these two equations
(multiplying the second equation by $-i$ and adding this to the
first) one finds
\begin{eqnarray}\label{holomorphic equation ising}
-e^{\frac{i \pi}{2}}s_{1}\mu_{\tilde{1}}-ie^{-\frac{i
\pi}{4}}s_{2}\mu_{\tilde{1}}+s_{2}\mu_{\tilde{2}}+ie^{\frac{i
\pi}{4}}s_{1}\mu_{\tilde{2}}=0\,.
\end{eqnarray}
Note that for this to happen  it is crucial that we use the
critical value $x_{c1}=\tan(\pi/8)=\sqrt2-1$. (For the special
case of the Ising model, when $x\not= x_c$ the right hand side of
(\ref{holomorphic equation ising}) is proportional to the
antiholomorphic fermion \cite{cardy}.)

(\ref{holomorphic equation ising}) has the form of a discrete
contour integral around each
 elementary square of the covering lattice (the union of the dual
 lattice with the original lattice)
\begin{eqnarray}\label{holomorphic equation}
\sum_{e}\psi_{e}\delta
z_{e}=0\qquad\mbox{with}\qquad\psi_{r\tilde{r}}=e^{\frac{-i\theta_{r\tilde{r}}}{2}}s_{r}\mu_{\tilde{r}}\,,
\end{eqnarray}
where $\theta_{r\tilde{r}}$ is the angle that the directed segment
$r\tilde{r}$ makes with the $x$-axes, with the convention $-\pi
\leq \theta_{r\tilde{r}}<\pi$, In this case these angles are just
$-\pi,-\frac{\pi}{2},0,\frac{\pi}{2}$ but in the next section we
will see more general cases. By changing $i\rightarrow -i$ one can
find the similar equation for the discrete antiholomorphic
fermions. These two linear equations are a lattice discretised
form of the Cauchy-Riemann equations.


The calculation for $N=3$ is very similar. Multiplying both sides
of equation (\ref{relation between disorder operators}) by the
denominator, and then by $s_{1}$, $s_{1}^{2}s_{2}^{*}$ and $s_{2}$
we can find three linear equations in the six variables
$s_{1}\mu_{\tilde1,\tilde2},s_{1}^{2}s_{2}^{*}\mu_{\tilde1,\tilde2},s_{2}\mu_{\tilde1,\tilde2}$.
By eliminating $s_{1}^{2}s_{2}^{*}\mu_{\tilde1,\tilde2}$ one again
finds a discretely holomorphic equation of the form
(\ref{holomorphic equation}) with now
\begin{eqnarray}\label{N=3}
\psi_{r\tilde{r}}=e^{\frac{-2i\theta_{r\tilde{r}}}{3}}s_{r}\mu_{\tilde{r}}\,.
\end{eqnarray}
Once again, this works only at the FZ point given by
(\ref{FZpoint}).

The case $N=4$ is very similar and we will discuss the results
later. $N=5$ is more interesting. This model has two FZ points
$(x_{1},x_{2})=(x_{c1},x_{c2})$ and $(x_{c2},x_{c1})$
corresponding to the transformation $s_r\to s_r^2$ which exchanges
$x_1$ and $x_2$. For this case the equation (\ref{relation between
disorder operators}) has the following form
\begin{eqnarray}\label{N=5}
&&\mu_{\tilde{1}m}(1+x_{c1}s_{1}s_{2}^{*}+x_{c2}(s_{1}s_{2}^{*})^{2}+
x_{c2}(s_{1}s_{2}^{*})^{3}+x_{c1}(s_{1}s_{2}^{*})^{4})\nonumber\\&&\qquad=
\mu_{\tilde{2}m}(1+e^{-\frac{2\pi i
m}{5}}x_{c1}s_{1}s_{2}^{*}+e^{-\frac{4\pi i
m}{5}}x_{c2}(s_{1}s_{2}^{*})^{2}+e^{-\frac{6\pi i
m}{5}}x_{c2}(s_{1}s_{2}^{*})^{3}+e^{-\frac{8\pi i
m}{5}}x_{c1}(s_{1}s_{2}^{*})^{4})\,.
\end{eqnarray}
For the $m=1$ case by multiplying the above equation by
$s_{1},s_{2},s_{1}^{2}s_{2}^{*},s_{1}^{3}s_{2}^{*2},s_{1}^{4}s_{2}^{*3}$
one finds the following equation
\begin{eqnarray}\label{N=5 m=1}
&&\mu_{\tilde{1}1}(s_{1}+x_{c1}s_{1}^{2}s_{2}^{*}+x_{c2}s_{1}^{3}s_{2}^{*2}
+x_{c2}s_{1}^{4}s_{2}^{*3}+x_{c1}s_{2}^{*4})\nonumber\\&&\qquad=
\mu_{\tilde{2}1}(s_{1}+e^{-\frac{2\pi i
}{5}}x_{c1}s_{1}^{2}s_{2}^{*}+e^{-\frac{4\pi i
}{5}}x_{c2}s_{1}^{3}s_{2}^{*2}+e^{-\frac{6\pi i
}{5}}x_{c2}s_{1}^{4}s_{2}^{*3}+e^{-\frac{8\pi i
}{5}}x_{c1}s_{2})\,,
\end{eqnarray}
and four other equations by permuting the terms
$s_{1},s_{1}^{2}s_{2}^{*},s_{1}^{3}s_{2}^{*2},s_{1}^{4}s_{2}^{*3},s_{2}$.
The next step is to eliminate the terms involving
$s_{1}^{2}s_{2}^{*},s_{1}^{3}s_{2}^{*2},s_{1}^{4}s_{2}^{*3}$ from
the above five equations. The calculation is rather cumbersome, so
we resorted to using Mathematica. The result is as follows: for
$m=1$ it is possible to eliminate the above terms only at the
first FZ point $(x_{c1},x_{c2})$, in which case we obtain the
holomorphic equation (\ref{holomorphic equation}) with the
holomorphic variable
$\psi^1_{r\tilde{r}}=e^{-\frac{4i\theta_{r\tilde{r}}}{5}}s_{r}\mu_{\tilde{r}1}$.

For $m=2$ we first multiply the equation (\ref{N=5}) by
$s_{1}^{2},s_{1}^{3}s_{2}^{*},s_{1}^{4}s_{2}^{*2},s_{2}^{*3},s_{1}s_{2}^{*4}$
to find
\begin{eqnarray}\label{N=5 m=2}
&&\mu_{\tilde{1}2}(s_{1}^{2}+x_{c1}s_{1}^{3}s_{2}^{*}+x_{c2}s_{1}^{4}s_{2}^{*2}
+x_{c2}s_{2}^{*3}+x_{c1}s_{1}s_{2}^{*4})\nonumber\\&&\qquad=
\mu_{\tilde{2}2}(s_{1}^{2}+e^{-\frac{4\pi i
}{5}}x_{c1}s_{1}^{3}s_{2}^{*}+e^{-\frac{8\pi i
}{5}}x_{c2}s_{1}^{4}s_{2}^{*2}+e^{-\frac{12\pi i
}{5}}x_{c2}s_{2}^{*3}+e^{-\frac{16\pi i
}{5}}x_{c1}s_{1}s_{2}^{*4})\,,
\end{eqnarray}
and four other equations by permuting the different terms. Again
we should eliminate the terms in involving variables
$s_{1}^{3}s_{2}^{*},s_{1}^{4}s_{2}^{*2},s_{1}s_{2}^{*4}$. In this
case we can find a holomorphic equation only at the second FZ
point $(x_{1},x_{2})=(x_{c2},x_{c1})$ with the holomorphic
variable
$\psi^2_{r\tilde{r}}=e^{\frac{-6i\theta_{r\tilde{r}}}{5}}s_{r}^{2}\mu_{\tilde{r}2}$.

The calculation for $m=3$ and $m=4$ is similar and gives the
antiholomorphic parafermions.

The generalization to higher values of $N$ is straightforward. By
simplifying equation (\ref{relation between disorder operators})
at the FZ points we conjecture that the following variables will
be discretely holomorphic:
\begin{eqnarray}\label{holomorphic spinors}
\psi^{m}_{r\tilde{r}}=e^{-ip_{m}\theta_{r\tilde{r}}}s^{m}_{r}\mu_{\tilde{r}m}\,,
\end{eqnarray}
for $1\leq m\leq[N/2]$. The same expression with the argument of
the exponential having the opposite sign is discretely
antiholomorphic for $[N/2]\leq m\leq N-1$. (This is just the
complex conjugate of $\psi^{N-m}_{r\tilde r}$.) We checked the
conjecture for $N=6$. In this case we find all of the holomorphic
operators at the one FZ point $(x_{c1},x_{c2},x_{c3})$.
These discretely holomorphic quantities are the candidates for the
holomorphic parafermions of FZ parafermionic CFT in the continuum
limit. We should emphasise that these operators are holomorphic
only at the FZ points and not elsewhere on the critical self -dual
surfaces. For example, for $N=4$, $\psi^{1}_{r\tilde{r}}$ is not
holomorphic at the critical point of the four-state Potts model.
However $\psi^{2}_{r\tilde{r}}$, which has conformal spin 1, is
holomorphic all the way along the critical line
$2x_{c1}+x_{c2}=1$.

We should also note another curious feature, which is most simply
illustrated for the Ising case. Although we have shown that the
linear relation (\ref{holomorphic equation ising}) implies the
discrete holomorphicity relation (\ref{holomorphic equation}) for
the parafermion $\psi_{r\tilde r}=e^{-\frac{i\theta_{r\tilde
r}}{2}}s_r\mu_{\tilde r}$, if we define another quantity
$\hat\psi_{r\tilde r}\equiv e^{\frac{3i\theta_{r\tilde
r}}{2}}s_r\mu_{\tilde r}$, then (\ref{holomorphic equation ising})
also implies that $\sum_e\hat\psi_e\delta z_e^*=0$, that is
$\hat\psi_{r\tilde r}$ is discretely antiholomorphic. This is
quite general, and holds for other values of $N$ and the more
general lattices discussed in the next section: if $\psi_{r\tilde
r}$ is discretely holomorphic, then $\hat\psi_{r\tilde
r}=e^{2i\theta_{r\tilde r}}\psi_{r\tilde r}$ is discretely
antiholomorphic.

\section{ Z$_N$ model on other lattices}
In the previous section we found additional evidence that the
Zamolodchikov-Fateev parafermionic CFT is a good candidate for the
continuum limit of the FZ point of Z$_N$ models on the square
lattice, by identifying discretely holomorphic parafermions on the
lattice. Note this construction was possible only at the FZ
points. One may therefore try to invert the argument and locate
the FZ point by requiring discrete holomorphicity. Since, by
universality, the same conformal field theory should describe the
continuum limit of suitable Z$_N$ models on other lattices, one
would expect to be able to identify discretely holomorphic objects
in this case, and thereby deduce the location of the FZ points. In
fact we will show that this is possible in many cases, and that
the critical weights of the models are just related to the
geometry of the covering lattice when it is suitably embedded in
the plane. Note that the notion of holomorphicity is not invariant
under a general diffeomorphism of the plane, only under conformal
transformations. Therefore we expect that the identification of
discretely holomorphic objects will depend on choosing a
particular embedding of the lattice into the plane, modulo
conformal mappings. In what follows we study this problem for the
anisotropic square, honeycomb and triangular lattices, and then
the more general case of a `Baxter lattice'.

\begin{figure}[h]
\centering
\includegraphics[width=6cm]{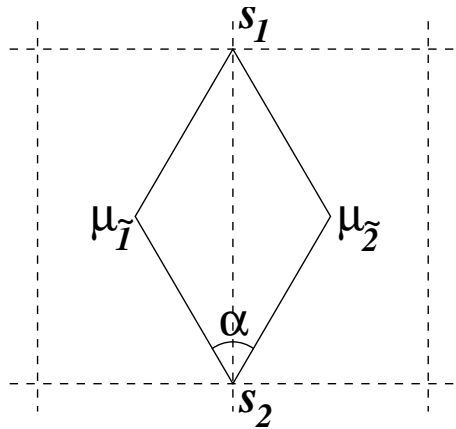}
\caption{\label{fig2} Embedding of the anisotropic square lattice
(dashed lines) in the plane. The faces of the covering lattice
form rhombi with opening angles $\alpha$ (shown) and $\pi-\alpha$.
}
\end{figure}

Let first investigate the square lattice with unequal weights
$x_k^x$, $x_k^y$ in the $x$ and $y$ directions respectively. In
this case it is clear that in order to maintain invariance under
lattice translations and reflection symmetry about the $x$ and $y$
axes the only transformations allowed are relative scalings of the
$x$ and $y$ coordinates \cite{barber}. In this case each
elementary square of the covering lattice is deformed into a
rhombus (Fig.~\ref{fig2}).

%
%


Defining the parafermions by (\ref{holomorphic spinors}), we can
try to demand that the discrete contour integral around each
elementary rhombus vanish as before. As an example we show the
case of the Ising model in detail. Taking an arbitrary linear
combination of (\ref{s equation}) and its counterpart with
$s_1\leftrightarrow s_2$ we have
\begin{equation}
\label{rhomb1}
\frac{a+x_1^y}{ax_1^y-1}s_1\mu_{\tilde1}+\frac{ax_1^y+1}{ax_1^y-1}s_2\mu_{\tilde1}
+\frac{a-x_1^y}{1-ax_1^y}s_1\mu_{\tilde2}+s_2\mu_{\tilde2}=0
\end{equation}
On the other hand, for the rhombus in Fig.~\ref{fig2} the contour
sum is
\begin{equation}
\label{rhomb2}
-\psi_{1\tilde1}+e^{-i(\pi-\alpha)}\psi_{2\tilde1}+\psi_{2\tilde2}+e^{i\alpha}\psi_{1\tilde2}=0
\end{equation}
where $\psi_{1\tilde1}=e^{i\pi/2}s_1\mu_{\tilde1}$,
$\psi_{2\tilde1}=e^{-i\alpha/2}s_2\mu_{\tilde1}$,
$\psi_{2\tilde2}=s_2\mu_{\tilde2}$ and
$\psi_{1\tilde2}=e^{i(\pi-\alpha)/2}s_1\mu_{\tilde2}$. Comparing
(\ref{rhomb1}) and (\ref{rhomb2}) we find that they are consistent
only at a particular value of $a$ and for $x_1^y=\tan(\alpha/4)$.
Similarly we find the contour sums around the rhombi aligned in
the other direction can vanish only if
$x_1^x=\tan((\pi-\alpha)/4)$.

Note that the comparison of (\ref{rhomb1},\ref{rhomb2}) implies 3
complex equations. If instead of choosing the elementary face of
the covering lattice to be a rhombus we take an arbitrary
quadrilateral with edges $\delta z_e$, where $\sum_{e=1}^4\delta
z_e=0$, this would introduce 2 further complex unknowns (ratios of
the $\delta z_e$) into (\ref{rhomb2}), besides the unknown $a$.
Given $x_1^y$, the 3 complex equations therefore determine $a$ and
the shape of the quadrilateral. We have already shown that a
rhombus with a suitable opening angle satisfies all the equations.
Therefore one would expect that they cannot be satisfied for any
other quadrilateral with unequal edge lengths. This can be checked
explicitly.

For higher values of $N$ we find that the contour sum around the
rhombi vanish as long as the weights satisfy
\begin{eqnarray}\label{FZpoint anisotropic case}
x^{y}_{k}&=&x_{ck}(\alpha)\equiv\prod_{i=0}^{k-1}\frac{\sin(\frac{\pi
i}{N}+\frac{\alpha}{2N})}{\sin(\frac{\pi
(i+1)}{N}-\frac{\alpha}{2N})}\nonumber\\x^{x}_{k}(\alpha)&=&x_{ck}(\pi-\alpha)\,.
\end{eqnarray}
The result agrees with that found by Fateev and Zamolodchikov
\cite{FZ star triangle} by imposing the the star-triangle
relations.

Next we consider the isotropic honeycomb and triangular lattices,
which are mutually dual with the vertices of the honeycomb lattice
at the centers of the faces of the triangular lattice and \em vice
versa\em. The elementary cells of the covering lattice give a
regular rhombus tiling of the plane, with $\alpha=\frac{2\pi}{3}$
for the honeycomb lattice and $\alpha=\frac{\pi}{3}$ for the
triangular lattice.  The mathematics then proceeds just as for the
anisotropic square lattice above, with the result that one can
find discretely holomorphic parafermions as long as the weights
satisfy (\ref{FZpoint anisotropic case}) with the appropriate
values of $\alpha$. This we then conjecture to correspond to the
FZ point(s) for these lattices.

\begin{figure}[h]
\centering
\includegraphics[width=8cm]{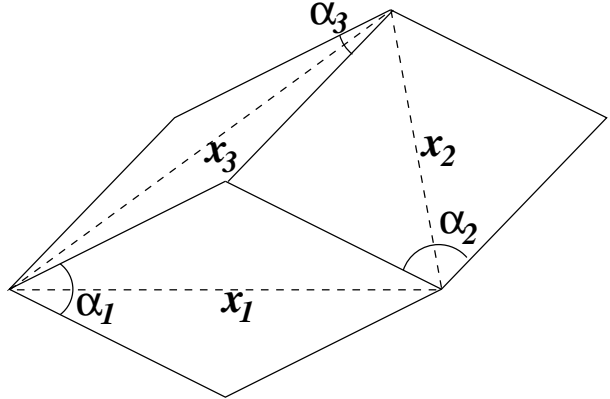}
\caption{\label{fig3} Embedding of one face of an regular
anisotropic triangular lattice (dashed lines) in the plane. The
faces of the covering lattice are rhombi with opening angles
$\alpha_1$, $\alpha_2$ and $\alpha_3=\pi-\alpha_1-\alpha_2$.}
\end{figure}

A more interesting case is that of the homogeneous triangular
lattice with unequal weights $x_k^{(1)}$, $x_k^{(2)}$ and
$x_k^{(3)}$ in the three lattice directions. In this case, the
question of where to locate the dual vertices is crucial. On the
basis of the above observations, we adopt the following
construction. Consider a (proper) embedding of the regular
triangular lattice into the plane by some general linear
transformation of the coordinates. This gives a regular tiling of
the plane by triangles. For each triangular face with Z$_N$ spins
at the vertices, locate the dual vertex at the \em circumcenter
\em, that is the point at which the three perpendicular bisectors
of the edges meet, equidistant from the three vertices. This
construction guarantees that adjacent pairs of vertices and dual
vertices always form a rhombus. Each triangle is associated with
three different rhombi with angles $\alpha_1$, $\alpha_2$, and
$\alpha_3=\pi-\alpha_1-\alpha_2$, see Fig.~\ref{fig3}. If now we
choose the weights so that
\begin{eqnarray}\label{FZpoint anisotropic triangular lattice}
x_{1k}=x_{ck}(\alpha_{1})\,,\hspace{2cm}x_{2k}=x_{ck}(\alpha_{2})\,,
\hspace{2cm}x_{3k}=x_{ck}(\pi-\alpha_{1}-\alpha_{2})\,,
\end{eqnarray}
it follows from our general analysis that we then can identify
discretely holomorphic parafermions. Transforming back to the
orginal regular triangular lattice with unequal weights, we
therefore conjecture that (\ref{FZpoint anisotropic triangular
lattice}) gives the general critical surface of the FZ point(s).
Note that it is possible for one of the $\alpha_j$ to be negative,
if the largest angle of the deformed triangle is obtuse,
corresponding to the interactions in one of the lattice directions
being (weakly) antiferromagnetic. (However these points are still
on the ferromagnetic critical surface, and have the
Fateev-Zamolodchikov CFT \cite{parafermion FZ} as their
conjectured scaling limit.) On transforming back to the regular
lattice, the images of the dual vertices do not in general lie at
the centers of the triangular faces (in fact they may lie outside
the face).

It is easy to check that these results are consistent with
previously known results for the $N=2$ Ising model and $N=3$ three
states Potts model \cite{baxter,wu}.









The FZ critical points for the honeycomb lattice with anisotropic
couplings follow directly by duality from those of the triangular
lattice, which corresponds to letting $\alpha_j\to\pi-\alpha_j$ in
(\ref{FZpoint anisotropic triangular lattice}).

\begin{figure}[h]
\centering
\includegraphics[width=8cm]{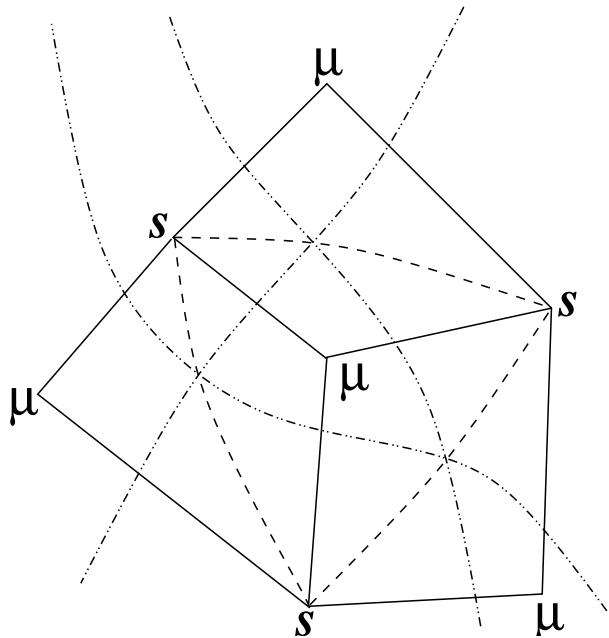}
\caption{\label{fig4} Part of a Baxter lattice. The faces of the
graph $\cal L$ formed by the curved lines are 2-colorable (not
shown). Order variables $s$ and disorder variables $\mu$ are
associated with alternately colored faces respectively. The
covering lattice is shown as solid lines. The theorem of Kenyon
and Schlenker \cite{kenyon} asserts that for every such graph the
covering lattice admits a rhombic embedding in the plane, that is
one where all its edges have the same length.}
\end{figure}

We now indicate how our results extend to a more general kind of
inhomogeneous lattice, called a Z-invariant or Baxter lattice
\cite{baxter}. This is a planar graph $\cal L$ which is a union of
$M$ simple (non-self-intersecting) curves crossing the complex
plane from $x_j-i\infty$ to $x_j'+i\infty$, where the values
$\{x_j\}$ and $\{x_j'\}$ ($1\leq j\leq M$) are distinct, and with
the further property that a given curve can intersect any of the
others at most once (Fig.~\ref{fig4}). The faces of $\cal L$ are
2-colorable, and in general we can assume that its vertices are
all of degree four (if not we deform the curves slightly so this
is true.) Consider now the planar graph $\cal G$ whose vertices
are associated to each black face of $\cal L$, and whose edges $E$
pass through the vertices of $\cal L$. We can define a Z$_N$ model
on $\cal G$ with general weights $x_k^{(E)}$. The vertices of the
dual lattice ${\cal G}^*$ then correspond to the white faces of
$\cal L$. The vertices of the covering graph $\cal C$ (the union
of the vertices of $\cal G$ and ${\cal G}^*$) each correspond to a
face of $\cal L$, irrespective of color. Note that all of the
faces of $\cal C$ have degree 4. We are free to embed this lattice
in the plane in any way we choose, just as for the anisotropic
triangular lattice earlier. However, a remarkable theorem due to
Kenyon and Schlenker \cite{kenyon} states that (in the case
considered here, where the original curves do not self-intersect
and cross any other at most once) there is a \em rhombic embedding
\em of $\cal C$ into the plane, that is, one in which all the
edges have equal length. Each rhombus corresponds to an edge $E$
of $\cal G$, and defines an opening angle $\alpha_E$. As before,
we may define parafermions on the edges of $\cal C$ and demand
that they be discretely holomorphic when summed around the edges
of each rhombus. This will be the case if we choose
$x_k(E)=x_{kc}(\alpha_E)$, as given by (\ref{FZpoint anisotropic
case}). We conjecture that these values give the location of the
FZ points on this inhomogeneous model. The rhombic embedding
specifies how this model should be embedded in the plane in order
that its scaling limit be given by the FZ conformal field theory.

\begin{figure}[h]
\centering
\includegraphics[width=12cm]{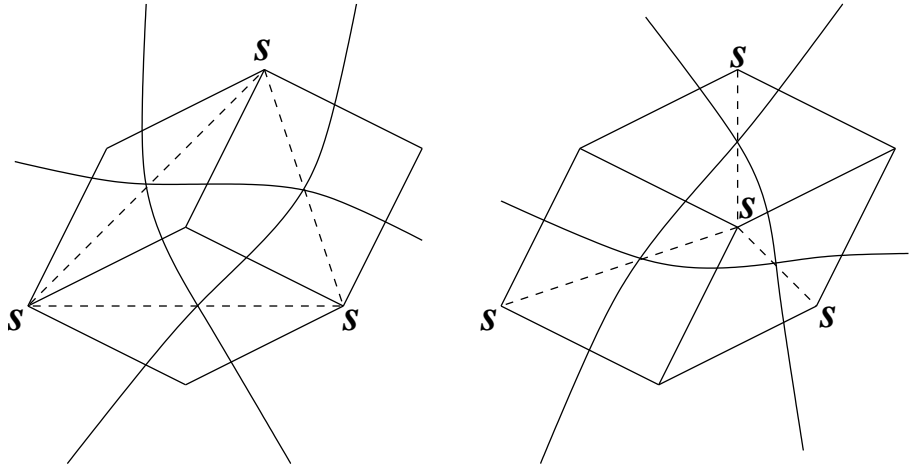}
\caption{\label{fig6} Two different tilings of a hexagon by the
same set of three rhombi. The right-hand case has, in the graph
$\cal G$, an additional vertex associated to an order operator $s$
as compared to that on the right. Discrete holomorphicity for each
rhombus fixes the couplings on the dashed lines to be related by
the star-triangle transformation. The two pictures are also
related in the original graph $\cal L$ by moving one of the curves
past the vertex formed by the other two -- the Yang-Baxter
relation.}
\end{figure}

Finally we discuss how the conditions that allow the Yang-Baxter
(star-triangle) relations in these models are compatible with
discrete holomorphicity. Consider part of a lattice $\cal L$ where
three lines meet at a point. This can be resolved in two different
ways (see Fig.~\ref{fig6}). The Yang-Baxter relations guarantee
that the Boltzmann weights, keeping all the other lines fixed, are
independent of how this is done. If we 2-color the faces of $\cal
L$, it can be seen that the second resolution adds one more black
face, that is one more vertex of $\cal G$. The invariance under
this is the star-triangle relation.

The rhombic embeddings in the two cases are shown in
Fig.~\ref{fig6}. It can be seen that in both cases the three
rhombuses fit together to form a hexagon with opposite sides
parallel, and that they simply correspond to two different tilings
of the hexagon by the same three rhombi. The condition that the
contour sum of the parafermions around each rhombus should vanish
implies that the couplings should satisfy $\tilde
x_k=x_{kc}(\pi-\alpha)$, which are the critical star-triangle
relations.

\section{ Conclusion and Discussion}
In this paper we considered the nearest neighbour Z$_N$ model on
some fairly general lattices. We identified the dual disorder
operators, and used these to define parafermionic operators which
reside on the edges of the covering lattice. For the isotropic
square, triangular and honeycomb lattices we showed that, at the
Fateev-Zamolodchikov critical points, these parafermions obey the
discrete version of the Cauchy-Riemann relations, that is their
contour sum around each elementary face of the covering lattice
vanishes. We then extended this idea to regular lattices with
anisotropic couplings, and showed that, if they are correctly
embedded in the plane, it is possible once again to identify
discretely holomorphic parafermions at particular values of the
couplings. These we conjecture to correspond to the FZ points on
these anisotropic lattice, which agrees with previously known
cases.

A crucial feature of these embeddings is that the faces of the
covering lattice should be rhombi. This enabled us to extend our
results to more general inhomogeneous lattices of the Baxter type,
thanks to a theorem \cite{kenyon} which guarantees the existence
of a rhombic embedding in such cases. In this picture, the
relation between discrete holomorphicity and the Yang-Baxter, or
star-triangle, relations becomes particularly clear.

We expect that the discretely holomorphic lattice parafermions
become the holomorphic parafermions of the conformal field theory,
but this requires further assumptions. As pointed out in
Refs.~\cite{smirnov,cardy}, if we regard the discrete
Cauchy-Riemann equations as a linear system, there are in general
twice as many unknowns as equations (only in the Ising case, when
the phases of the parafermions are not free, are there the correct
number). Morera's theorem, which allows one to deduce that a
function $f$ defined on ${\mathbb R}^2$ is analytic if its contour
integral around every closed contour vanishes, applies only if $f$
is also assumed to be continuous. If we could prove that all
correlators of our discrete parafermions become continuous
functions in the scaling limit, this would be sufficient to show
complex analyticity, at least with sufficiently smooth boundaries.
However this step is highly non-trivial, as evidenced by the
ambiguity in the interpretation of the linear relations discussed
at the end of Sec.~\ref{sec2}. There we showed that they lead to
the identification of discrete parafermions $\psi_{r\tilde r}$ and
also $\hat\psi_{r\tilde r}=e^{2i\theta_{r\tilde r}}\psi_{r\tilde
r}$. If they corresponded to conformal fields in the scaling
limit, they would have conformal spins $p_m$ and $p_m-2$. However,
if the correlators of $\psi$ become continuous functions in the
scaling limit, this cannot be true of those of $\hat\psi$, and
vice versa.

If this problem can be overcome, we believe that our
identification of suitable discretely holomorphic quantities
should be the first step in showing that suitable defined curves
in the Z$_N$ have SLE as their scaling limit, as has been recently
conjectured \cite{santachiara,gamsacardy}. In order to do this,
following the ideas of Smirnov \cite{smirnov}, however, it is
necessary to identify these quantities with observables of these
curves which are martingales of some discrete exploration process.
Since the domain walls (or, equivalently, high-temperature graphs)
of the Z$_N$ model do not correspond directly to simple lattice
curves (except for $N=2$), there are a number of difficulties yet
to be overcome in this program.

Acknowledgements:  J.C.~thanks V.~Riva for early discussions of
this problem, and acknowledges partial support from EPSRC grant
EP/D050952/1. M.R.~is grateful to the British Council, Iran and
the Center of Excellence in Physics, Sharif University of
Technology, Iran for financial support to visit the University of
Oxford. This work was carried out when M.R. was a visitor at the
Rudolf Peierls Centre for Theoretical Physics, University of
Oxford, and he expresses his thanks for support and hospitality
during the visit. Finally, special thanks go to F.~Ahmadi who
solved many of the problems in obtaining financial support and
visa.

\end{document}